\documentclass[a4paper,11pt,oneside]{amsart}
\usepackage[T1]{fontenc}              
\usepackage{rotating}
\usepackage{lmodern} 
\usepackage{amsmath,amssymb,amscd,bm,mathtools,xypic,extpfeil,graphicx,color,transparent}   
\usepackage{amsthm}
\newtheorem{thm}{Theorem}
\theoremstyle{definition}
\newtheorem{defn}[thm]{Definition}
\theoremstyle{example}

\theoremstyle{remark}
\newtheorem{remark}[thm]{Remark}
\theoremstyle{lemma}
\newtheorem{lemma}[thm]{Lemma}
\theoremstyle{corollary}

\newtheorem{proposition}[thm]{Proposition}

\newcommand {\NN}{{\mathbb N}}
\newcommand {\QQ}{{\mathbb Q}}
\newcommand {\CC}{{\mathbb C}}
\newcommand {\RR}{{\mathbb R}}
\newcommand {\ZZ}{{\mathbb Z}}
\newcommand {\TT}{{\mathbb T}}
\newcommand {\DD}{{\mathbb D}}
\newcommand {\PP}{{\mathbb P}}
\newcommand {\KK}{{\mathbb K}}

\newcommand{\FF}{{\mathbb F}}
\newcommand{\HH}{{\mathcal H}^\alpha}
\newcommand{\HHprime}{{\mathcal H}^{\alpha'}}

\newcommand{\HHcz}{{\mathcal H}^{\check\alpha}}
\newcommand{\vN}{\mathcal{N}}
\newcommand{\pt}{\{ {\rm pt}\}}
\newcommand{\vertiso}{\begin{turn}{270}$\!\!\cong \;\;$\end{turn}}

\usepackage{amsmath}
\usepackage[all,cmtip]{xy}

\begin{document}
\title{Pairs of pants, Pochhammer curves and $L^2$-invariants}
\author{Marcel B\"okstedt}
\address{Institut for Matematik, Aarhus Universitet, Ny Munkegade 118, bygn.~1530, 8000 \AA rhus C, Denmark}
\email{marcel@imf.au.dk}
\author{Nuno M.~Rom\~ao}
\address{Mathematisches Institut, Georg-August-Universit\"at G\"ottingen, Bunsenstra\ss e 3--5, 37073 G\"ottingen, Germany}
\email{nromao@uni-math.gwdg.de}
\date{October 20, 2014}
\begin{abstract} We propose an intuitive interpretation for nontrivial $L^2$-Betti numbers of compact Riemann surfaces in terms of certain  loops in embedded pairs of pants. This description uses twisted homology associated to the Hurewicz map of the surface, and it satisfies a sewing property with respect to a large class of pair-of-pants decompositions. Applications to supersymmetric quantum mechanics incorporating Aharonov--Bohm phases are briefly discussed, for both point particles and topological
solitons (Abelian and non-Abelian vortices) in two dimensions.
\end{abstract}
\maketitle

\section{Introduction}

The theory of $L^2$-invariants was created out of an effort by Atiyah to extend the index theorem for elliptic operators on compact manifolds  to certain noncompact
situations. In a ground-breaking paper~\cite{AtiEO}, he proposed a {\em $\Gamma$-index} for elliptic differential operators on an infinite Galois cover $\tilde M$ of a compact  manifold $M$ which
combined analysis of operators over a fundamental domain for the action  of the group of deck transformations $\Gamma$, identified with an open dense subset of $M$, with analysis on the Hilbert space of sequences $\ell^2(\Gamma)$.
This machinery, applied to the operator ${\rm d} + {\rm d}^*$ on the manifold $\tilde M$ with a cocompact $\Gamma$-invariant Riemannian metric, leads to invariants $b^i_{(2)}$ or $b_i^{(2)}$  which generalize the rank of $L^2$-cohomology groups of finite covers; it can be used to great advantage in cases where the $L^2$-cohomology of the covering space is an infinitely generated vector space. These {\em $L^2$-Betti numbers} are obtained as Murray--von Neumann (or renormalized)
dimensions~\cite{MurvNeu} of certain Hilbert ${\vN}(\Gamma)$-modules, where ${\vN(\Gamma)}:={\mathcal B}(\ell^2(\Gamma))^\Gamma$ is the von Neumann algebra
associated to the discrete group $\Gamma$.  For example, if $M=\TT^n$ is an $n$-torus (the setting for classical Fourier analysis), applying this construction to the universal cover $\tilde M=\mathbb{R}^n$ leads to all $L^2$-Betti numbers $b^i_{(2)}(\RR^n,\Gamma)$ for  $\Gamma=\pi_1(\TT^n)\cong \ZZ^n$ being zero --- this follows from multiplicity under finite covers and a K\"unneth-type formula. Quite often,
$L^2$-Betti numbers vanish in situations where the ordinary
Betti numbers of the quotient space do not, and they are not necessarily integral real numbers; a question that was open until recently was whether they can ever be irrational. For a layout of the theory of $L^2$-invariants, and in order to appreciate the impact that these ideas have had in mathematics so far, we refer the reader to the textbook~\cite{Lue}. An informal account of some basic notions, tailored to the purposes of the present paper, will be given in section~\ref{secL2Bettis} below.

An example leading to nontrivial $L^2$-invariants is provided by the universal cover $\DD$ (the disc) of a compact oriented Riemann surface $M=\Sigma$ of genus $g>1$. This simple but somewhat crucial example was already considered in Atiyah's original paper, where he illustrated his new theory by showing that
\begin{equation} \label{Atiyah}
b^i_{(2)}(\tilde \Sigma,\Gamma)=\left\{ 
\begin{array}{ll}
0 & \text{ if $i\ne 1$},\\
2g-2 &  \text{ if $i = 1$}.
\end{array}
\right.
\end{equation}
Availing oneself of the standard machinery~\cite{Lue}, the argument  runs along the following lines. The $L^2$-Betti numbers in degrees $i=0$ or $2$ must vanish because $\DD$ has
no compact component, thus any nontrivial $L^2$-Betti numbers must
lie in degree 1. Then one uses that the $L^2$-Euler characteristic of a co-compact space equals the ordinary  Euler characteristic of the quotient to conclude that
\begin{equation} \label{L2Betti1}
b^1_{(2)}(\tilde \Sigma,\Gamma) =-\sum_{i=0}^2 (-1)^i b^i_{(2)}(\tilde \Sigma, \Gamma)=:-\chi_{(2)}(\tilde \Sigma, \Gamma)=-\chi(\Sigma)=2g-2.
\end{equation} 
As elegant as this argument may be, it throws very little light on the meaning attached to the result~(\ref{Atiyah}). In addition, the
definitions that have been provided for these invariants (three different versions are proposed in~\cite{Lue}, all of which coincide in Atiyah's example) do not seem to
give a  hint to their geometrical content. This is in contrast with ordinary Betti numbers in singular homology or de Rham cohomology, for instance.

In this paper, we attempt to remedy this situation, and place the formula (\ref{Atiyah}) on more intuitive ground. More precisely, our immediate interest is in 
the case where $\tilde \Sigma$ is not the universal cover of  $\Sigma$, but rather the maximal Abelian cover of the surface with the action of the Abelianization $H_1(\Sigma;\ZZ)$ of its fundamental group, which we take as $\Gamma$ instead of $\pi_1(\Sigma)$. This is
a somewhat easier version of Atiyah's example (admitting a slick proof that coincides with the one we have just sketched for the universal cover), and it is motivated
by certain questions in mathematical physics that we shall elaborate on in section~\ref{physics}. Concretely, our main goal is to interpret the formula (\ref{L2Betti1}) for the Abelian cover in terms of a set of $2g-2$ objects
naturally associated to pair-of-pants decompositions of $\Sigma$ --- in a similar spirit to textbook presentations of concrete bases for $H_1(\Sigma;\ZZ)\cong \ZZ^{2g}$ in singular
homology, or  $H^1(\Sigma;\RR)\cong \RR^{2g}$ in de Rham cohomology (see e.g.~\cite{FulAT} for the latter). The relevant objects to consider will emerge as certain equivariant homology classes that we will associate to Pochhammer 1-cycles on pairs of pants, as explained in section~\ref{seclemniscates}.
We shall  illustrate how this sort of interpretation plays a role in the context of gauge theory of charged particles on a surface $\Sigma$ in section~\ref{physics}, after connecting
$L^2$-Betti numbers to Witten's supersymmetric quantum mechanics~\cite{WitSMT} coupled to local systems on manifolds.

\section{Twisted homology of $\alpha$-covers} \label{secalpha}

We start by fixing an Abelian group $A\cong \ZZ^n$ of rank $n \in \mathbb{N}$. Let $R:=\ZZ[A]$ be the integral group ring of $A$, and let us denote by $\FF:={\rm Quot}(R)$ its field of fractions. In what follows, we shall be interested in the topology of a given space $X$ equipped with a group homomorphism 
\begin{equation}\label{alpha}
\alpha:\pi_1(X) \to A.
\end{equation}  
We may alternatively interpret
$\alpha$ in terms of a cohomology class $\bar \alpha \in H^1(X;A)$ via the universal coefficient theorem in cohomology~\cite{Hat}, since (\ref{alpha}) factors through 
the Hurewicz map. A natural choice might be to take $\alpha$ as the Hurewicz map itself modulo torsion and $n=b_1:={\rm rk}\,  H_1(X;\ZZ)$, but we shall
also be interested in the general situation.

There is a cover $p_\alpha: \tilde X \to X$ associated to each  $\alpha$ (or to its cohomological avatar $\bar\alpha$) such that $C_*(\tilde X)$
is a free $R$-module; we will refer to $p_\alpha$ and $\tilde X$ interchangeably as the {\em $\alpha$-cover} of $X$. 
To understand $p_\alpha$ in concrete terms, one could consider a classifying map
$\tilde \alpha:X \to {\rm B}A \cong (S^1)^n$ representing $\bar\alpha$, and
define $\tilde X$ to be the pull-back of the universal cover of the torus
$\RR^n \to (S^1)^n$ under this map $\tilde \alpha$. Note that the covering space $\tilde X$ of an $\alpha$-cover is not necessarily connected.
We can tensor the coefficients of the chain complex $C_*(\tilde X)$ with an $A$-module $M$ to obtain the $\pi_1(X)$-equivariant homology via the
homomorphism $\alpha$. This construction
provides us with homology groups enriched with information pertaining to the homotopy theory of $X$. For a judicious choice of $\alpha$, they have the potential of capturing information on the topology of $X$ that is missed out by the non-equivariant (e.g.~singular) homology or cohomology groups of the space $X$.
 
In this paper, we want to explore a context where the field $\FF$  provides an interesting choice of coefficients. Thus for $i \in \NN_0$ we shall consider the {\em $\alpha$-twisted homology groups} 
$$\HH_i(X):= H_i(C_*(\tilde X)\otimes_R\FF).$$
Observe that, for each $i$, there is an obvious map $\rho_i$ given by the composition
\begin{equation} \label{rho}
\rho_i : H_i(\tilde X;\ZZ) = H_i(C_*(\tilde X)) \cong H_i(C_*(\tilde X)\otimes_R R)\to H_i(C_*(\tilde X)\otimes_R \FF)=\HH_i(X).
\end{equation}
More generally, if $\mathbb{K}$ is a field of characteristic zero, we write $\FF_\KK:= {\rm Quot}( \KK [A]) \supset \FF$ and define
$\HH_i(X;\FF_\KK):=H_i(C_*(\tilde X) \otimes _R \FF_\KK)$. Since $\FF_\KK$ is necessarily projective over $\FF$, one has 
$\HH_i(X;\FF_\KK)\cong \HH_i(X)\otimes_\FF \FF_\KK$.
We shall sometimes refer to the $\alpha$-cover associated to the Hurewicz map of $X$ as its Hurewicz cover. 

If the homomorphism $\alpha$  is trivial (i.e.~it maps to $0\in A$), then the corresponding $\alpha$-cover  is simply
the second-component projection $A \times X \to X$. In this case, 
$C_*(\tilde X) \cong C_*(X)\otimes_\ZZ R$ and
$C_*(\tilde X)\otimes_R \FF
\cong C_*(\tilde X)\otimes_\ZZ R \otimes_R \FF  
\cong C_*(X) \otimes_\ZZ \FF$.
Thus for $\alpha=0$ we simply recover $\HH_i(X)\cong H_i(X;\FF)$, the homology of $X$ with 
{\em untwisted} coefficients in $\FF$. By the universal coefficient theorem in homology~\cite{Hat},  this equals $H_i(X;\QQ)\otimes_\QQ \FF$. 

More generally, suppose that $i': A'\cong \ZZ^{n'}\to A $ is an injective group homomorphism for some $0 \le n'\le n$, and that $\alpha$
factors as 
\begin{equation} \label{alphafactor}
\alpha: \pi_1(X) \xrightarrow{\alpha'} A' \xrightarrow{i'} A.
\end{equation}
Write $R' := \ZZ[A']$ and $\FF':={\rm Quot}(R')$; then $i'$ induces a field extension $\FF'\hookrightarrow  \FF$. 
If $\alpha$ is trivial, we can choose $\FF'$ to be the prime field $\QQ$, but
we will also be interested in the case when $\FF'$ is a field intermediate between $\QQ$ and $\FF$.

Let $\tilde X'$ denote the corresponding $\alpha'$-cover of $X$.
We have
\[
C_*(\tilde X)\otimes_R \FF \cong 
(C_*(\tilde X') \otimes_{R'} R) \otimes _R \FF
\cong C_*(\tilde X)\otimes_{R'}\FF' \otimes_{\FF'} \FF.
\]
Taking the homology of this complex, we obtain the following {\em reduction property}:
\begin{equation} \label{redprop}
\HH_*(X) = {\mathcal H}^{\alpha'}_*(X)\otimes_{\FF'}\FF.
\end{equation}

We will now look at special cases. The first one is $X=S^1$.
\begin{lemma} \label{HHS1}
For all $i \in \NN_0$,
\[
\HH_i(S^1) \cong
\left\{
\begin{array}{ll}
 H_i(S^1;\QQ)\otimes_\QQ\FF  & \text{if $\alpha$ is trivial},\\
 0 & \text{otherwise}.
\end{array}
\right. 
\]
\begin{proof}
The first case follows directly from the reduction argument above, while the second statement is a computation that we need to carry out. 
  
 Consider first the case 
$A=\ZZ$ and $\alpha={\rm id}_\ZZ$. 
Then $\widetilde{S^1} \cong \RR$ is the universal cover of the circle with the translation action of $\ZZ$, and  
$C_*(\widetilde{S^1})$ is the $A$-equivariant chain complex $C_1\cong R$,
$C_0\cong R$. The boundary map is multiplication by $1-[\gamma]$, where $\gamma$
generates $A$. Since this element of $R\subset \FF$ is nontrivial, after tensoring with
$\FF$ it induces an isomorphism, and we get that $\HH_*(S^1) \cong 0$.

The next case is $A=\ZZ$ and $\alpha$ nontrivial, say the image of $\alpha$
is the subgroup $A'=mA\subset A$ for some $m\ge 2$. We have that 
$\alpha$ factors as $\pi_1(X) \xrightarrow{\alpha'} A' \subset A$.
There is a corresponding field extension $\FF'\subset \FF$, and by the
reduction property, we have that
$H^{\alpha}_*(X) \cong  H^{\alpha'}_*(X)\otimes_{\FF'}\FF$, which proves the lemma in this case.
  
Finally, if $\alpha:\pi_1(S^1)\to A$ is any nontrivial homomorphism, its image is
contained in some split summand $A'\cong \ZZ$. Applying the previous case, we obtain that
${\mathcal H}_*^{\alpha'}(S^1)\cong 0$, and using the reduction property (\ref{redprop}) we obtain the
result.
\end{proof}
\end{lemma}

Suppose that we have a Mayer--Vietoris situation for the union $X = X'\cup X''$. A homomorphism
$\alpha:\pi_1(X) \to A$ restricts to homomorphisms from $\pi_1(X')$ and  $\pi_1(X'')$. 
Note that the choice of basepoints is immaterial, since $A$ is Abelian.
There is a a long exact sequence
\[
\cdots \to \HH_i(X'\cap X'') \to \HH_i(X')\oplus \HH_i(X'') \to \HH_i(X) \to \HH_{i-1}(X'\cap X'') \to \cdots
\]
In particular, if we know that the homology of the intersection $X'\cap X''$ is trivial, $\HH_i(X) \cong \HH_i(X') \oplus \HH_i(X'')$.  This {\em additivity property} underlies
the main application of $\alpha$-twisted homology in this paper, which we will make precise in section~\ref{seclemniscates}.

The next case we shall consider is the wedge product of two circles:  $X=S^1\vee S^1$. 
\begin{thm}
\label{homology.wedge}
  Supppose that the homomorphism $\alpha:\pi_1(S^1\vee S^1) \to A$ is nontrivial. Then
\[
\HH_i(S^1\vee S^1)=
\begin{cases}
  \FF &\text{ if $i=1$,}\\
0 & \text{ otherwise.}
\end{cases}
\]
Moreover, if $\alpha$ is trivial on one of the wedge factors, the inclusion of this factor induces 
an isomorphism on $\HH_1$.
\begin{proof}
 Let us label the two wedge factors $S^1_j$ by $j\in \{ 1,2 \}$. 
  First consider the case when $\alpha$ induces a nontrivial homomorphism on each wedge factor.
Then $\HH_*(S^1_j)=0$ for $j=1,2$, so, as a first application of the additivity property, the Mayer--Vietoris long exact sequence degenerates to an isomorphism $\HH_1(S^1_1\vee S^1_2) \to \HH_0(\pt)\cong \FF$. 

Since $\alpha$ is nontrivial, it cannot be nontrivial on both wedge components.  Assume without
loss of generality that it is nontrivial on $S^1_2$. Then $\HH_*(S^1_2)=0$, and the Mayer--Vietoris long exact sequence takes the form
\[
0\to \HH_1(S^1_1) \to \HH_1(S^1_1\vee S^1_2) \to \HH_0(\pt) \to \HH_0(S^1_1) \to \HH_0(S^1_1\vee S^1_2) \to 0.
\]
By reduction and naturality, since the map $H_0(\{{\rm pt}\};\QQ)\to H_0(S^1_1;\QQ)$ is an isomorphism, the map $\HH_0(\pt) \to \HH_0(S^1)$ is also an isomorphism.
It follows that $\HH_0(S^1_1\vee S^1_2)\cong 0$, and that the map $\HH_1(S^1_1) \to \HH_1(S^1_1\vee S^1_2)$ is an isomorphism. Since we know from Lemma~\ref{HHS1} that $\HH_1(S^1_1)\cong \FF$, this completes the proof of the theorem.
\end{proof}
\end{thm}

Of course it can happen that neither of the wedge factors evaluate trivially.

\begin{lemma}
\label{injectivity}
  The natural map 
$\rho_1:H_1(\widetilde{S^1_1\vee S^1_2};\ZZ)\to
\HH_1(S^1_1\vee S^1_2)$ defined by (\ref{rho}) is injective.
\begin{proof}
First assume that both of the fundamental classes of $S_1^1$ and $S_2^1$ 
evaluate nontrivially under the map $\alpha:\pi_1(S^1_1\vee S^1_2)\to A$.
  We build $S^1\vee S^1$ using the following pushout square:
\[
\begin{CD}
  \{x_1\}\amalg  \{x_2\} @>>> S^1_1 \amalg S^1_2 \\
@VVV        @VVV    \\
 \{x\} @>>> S^1_1 \vee S^1_2 
\end{CD}
\]
This induces Mayer--Vietoris sequences for both
$H_*(\widetilde{S^1_1\vee S^1_2};\ZZ)$ and $\HH_*(S^1_1\vee S^1_2)$, and a natural transformation
$\rho_*$ between them. It follows that the pushout diagram above 
gives a ladder whose upper row is given by the functors $X \mapsto H_*(\widetilde{X};\ZZ)$ and whose lower row is
given by the functors $X \mapsto \HH_*(X)$.
 The condition on $\alpha$ implies that 
$H_1(\widetilde{S_i};\ZZ)\cong 0$. Using this fact, we see that the ladder  includes the   
following diagram with exact upper row:
\[
\begin{CD}
0 @>>> H_1(\widetilde{S^1_1\vee S^1_2};\ZZ) 
@>>> H_0(\widetilde{\{x_1\}}\amalg  \widetilde{\{x_2\}};\ZZ) \\
@. @V\rho VV @V\rho VV \\
@. \HH_1(S^1_1\vee S^1_2)  @>>> \HH_0(\{x_1\}\amalg  \{x_2\})  
\end{CD}
\]
To prove the lemma it suffices to prove that 
$\rho_0:H_0(\widetilde{\{p_i\}};\ZZ) \to \HH_0(\{p_i\})$ is injective.
But this map can be identified with the inclusion of
$R=\ZZ[A]$ into its quotient field.

We also consider the case where one of the fundamental classes $[S^1_i]$ evaluates
trivially under $\alpha$. Lets assume $\alpha([S^1_1])=0$ and 
$\alpha([S^1_2])\not=0$. There is a homotopy automorphism $h$ of $S^1_1\vee S^1_2$ such that
$h_*[S^1_1]=[S^1_1]+[S^1_2]$ and $h_*[S^1_2]=[S^1_2]$. Apply the previous case to the
composed map $\alpha \circ h_*:\pi_1(S_1^1\vee S_2^1)\to A$. Finally, use naturality to conclude that
the lemma is also true for an $\alpha$ of this sort.   
\end{proof}
\end{lemma}

Let us fix a homorphism $\alpha: \pi_1(S^1 \vee S^1)\rightarrow A$. Suppose that $f:S^1\to S^1\vee S^1$ is a map such that the composition $\check\alpha:=\alpha \circ \pi_1(f):\pi_1(S^1) \to A$ is the trivial homomorphism. Then $f$ induces a map 
\begin{equation} \label{fstar}
f_*: \; \FF \cong \HHcz_1(S^1) \longrightarrow \HH_1(S^1\vee S^1)\cong \FF.
\end{equation}

The following statement is a generalization of the last part of Theorem~\ref{homology.wedge}:

\begin{lemma}
\label{representant}
Let $p: \widetilde{S_1^1\vee S_2^1} \to S_1^1\vee S_2^1$
be the covering corresponding to $\alpha$.
Let $f:S^1 \to \widetilde{S_1^1\vee S_2^1}$ a map which is nontrivial in homology. Then $(p\circ f)_*^\alpha:\HH_1(S^1) \to \HH_1(S_1^1\vee S_2^1)$ is an isomorphism.
  \begin{proof}
   $f_*:\HH_1(S^1)\cong \FF$, so $f_*$ is an isomorphism if and only if it is nontrivial. 
Consider the diagram
\[
\begin{CD}
 H_1(\widetilde{S^1};\ZZ) @>{f_*}>> H_1(\widetilde{S^1_1\vee S^1_2};\ZZ)\\
@V\rho VV @V\rho VV \\
  \HH_1(S^1) @>{f_*^\alpha}>> \HH_1(S^1_1\vee S^1_2)\\
\end{CD}
\]
By  Lemma~\ref{injectivity}, the right vertical map is injective. 
Since $f_*([S^1])\not=0$, it follows that $(p\circ f)_*([S^1])\not=0$.
\end{proof}
\end{lemma}

\begin{remark}
\label{homotopy.lemniscates}
If for instance $\alpha$ is not injective, consider any closed curve
$\gamma:S^1\to  S^1_1\vee S^1_2$ representing a nontrivial homology class in the
kernel of $\alpha$. By covering theory, it can be written as 
$p\circ f$ for some $f:S^1 \to \widetilde{S^1_1\vee S^1_2}$. Since
$(p\circ f)_*([S^1])\not =0$, certainly $f_*([S^1]\not =0$. By 
Lemma~\ref{representant},
$\gamma$ induces an isomorphism on $\HH_1$.

If $\alpha$ is injective, a commutator of the inclusion of the two circles will be a map $\gamma:S^1\to S^1_1\vee S^1_2$ which lifts to 
$f:S^1 \to \widetilde{S^1_1\vee S^1_2}$. It is easy to check that $f$ is not
trivial on homology, so this map $\gamma$ induces an isomorphism on $\HH_1$.
\end{remark}

Our last result in this section is a K\"unneth-type formula for twisted homology that will be needed in section~\ref{vorticesGLSM}.
Let $\KK\subset \CC$ be a subfield, for instance  $\KK=\QQ$ or $\KK=\CC$, and $\FF_\KK := {\rm Quot} (\KK[A])$ as before.

\begin{thm}[K\"unneth formula] \label{thmKuenneth}

Let $p: X \times Y \rightarrow Y$ be the projection,
$\alpha: \pi_1(Y) \rightarrow A$ and $\alpha' := \alpha \circ \pi_1(p)$. Then 
$$ \HHprime_i(X\times Y;\FF_\KK)\cong \bigoplus_{j+k=i}  H_j(X;\QQ) \otimes_{\QQ} \HH_k(Y;\FF_\KK) .$$
\begin{proof}
  The Alexander--Whitney chain homotopy equivalence $C_*(\tilde X \times \tilde Y)\to C_*(\tilde X) \otimes_\ZZ C_*(\tilde Y)$ is natural for pairs of continuous maps. It follows that it is equivariant under $\ZZ[\pi_1(X\times Y)]\cong \ZZ[\pi_1(X)\times \pi_1(Y)]$, so that
it induces a chain homotopy equivalence
\begin{align*}
C_*(X\times Y)\otimes_{\ZZ[\pi_1(X\times Y)]}\FF_\KK &\to
C_*(\tilde X) \otimes_\ZZ C_*(\tilde Y)\otimes_{\ZZ[\pi_1(X)]\otimes \ZZ[\pi_1( Y))]}\FF_\KK.
\end{align*}
Since $\pi_1(X)$ acts trivially on $\FF_\KK$, this last complex equals
$C_*(X) \otimes_\ZZ C_*(\tilde Y)\otimes_{\ZZ[\pi_1( Y))]}\FF_\KK$,
which again equals
$(C_*(X)\otimes_\ZZ\QQ)\otimes_\QQ (C_*(\tilde Y)\otimes_{\ZZ[\pi_1( Y))]}\FF_\KK)$.
Now apply the usual K\"u{}nneth formula to this complex.
\end{proof}
\end{thm}

\section{Pochhammer principle for pair-of-pants decompositions} \label{seclemniscates}

In this section, we will take $\Sigma$ to be any oriented surface with negative Euler characteristic, which we assume to be compact (possibly with boundary). Topologically, we can think of it as a closed surface 
of genus $g$ with $h$ holes or punctures (realized by removing open discs), with the condition $$\chi(\Sigma)=2-2g-h<0$$
being enforced. In later sections, we shall specialize to the closed case where $h=0$, but not for the time being.

On such a surface $\Sigma$, we shall consider pair-of-pants decompositions $\{P_j\}_j$. Each $P_j$, which we take to be open, is homeomorphic to a 2-sphere with three closed discs removed, and the boundary $\partial P_j \subset \Sigma$ is a disjoint union of circles.
Recall that the Hurewicz cover $p_\alpha$ of $\Sigma$ is associated to the
Abelianization homomorphism
 $$\alpha:\pi_1(\Sigma)\to A=:H_1(\Sigma;\ZZ).$$ 
We can restrict this $\alpha$ to any pair of pants of the decomposition, and also to any of its boundary circles, but of course the corresponding covers will in general not be Hurewicz for those subsets. We say that a particular pair of pants $P \subset \Sigma$ is {\em fashionable} if all of its boundary circles $S^1_k \subset \partial P$  (there are at most three such circles) define  nontrivial elements of the 1-homology group $H_1(  \Sigma;\ZZ)$; and that a pair-of-pants decomposition $\{P_j\}_j$
of $\Sigma$ is fashionable if all its $P_j$ are.
For instance, the pair of pants $P\subset \Sigma$ illustrated in Fig.~2 is fashionable, but the pairs of pants $P_1$ and $P_2$ in Fig.~3 are not.

The following result justifies considering  fashionable pair-of-pants decompositions.

\begin{thm} \label{whyfashion}
\label{decomposition}
Let $\Sigma$ be an orientable surface with $\chi(\Sigma)=2-2g-h<0$. Given a fashionable pair-of-pants decomposition $\{P_j\}_j$ of $\Sigma$, the 
inclusions  $i_j:P_j \hookrightarrow \Sigma$ induce an isomorphism
\[
\FF^{2g-2+h} \cong \bigoplus_{j} \HH_1(P_j)\stackrel{\cong}{\longrightarrow} \HH_1(\Sigma).
\]
\begin{proof}
If $d:=2g-2+h=1$, then $\Sigma$ is either a 2-sphere with three punctures (homeomorphic to the closure of a pair of pants) or a 2-torus with one puncture (the closure of a pair of pants with two of its circle boundaries
identified). In the former case there is nothing to show, whereas in the latter case we observe that the interior circle $S^1_0$ formed by the identification of the two boundary components
is such that $\HH_1(S^1_0)\cong 0$ by Lemma~\ref{HHS1}, and the statement follows.

We claim inductively (on $d>1$) that, if $X\subset \Sigma$ is the closure in $\Sigma$ of 
a union of pairs of pants in our decomposition, we have an isomorphism
 \[
\bigoplus_{P_j \subset X} \HH_1(P_j)\stackrel{\cong}{\longrightarrow}  \HH_1(X).
\]
This follows from the induction hypothesis, and from the assumption that each $P_j$ is fashionable:  for each boundary circle $S^1_{j,k}\subset \bar P_j$, Lemma~\ref{HHS1}
again implies that $\HH_*(S^1_{j,k})\cong 0$. The well-known fact that any pair-of-pants decomposition of a surface of genus $g$ with $h$ punctures has exactly $d=2g-2+h$ elements is a consequence of
 the additivity property of $\chi$ under glueing and $\chi(P_j)=-1$, $\chi(S^1)=0$, $
 \chi(\DD)=1$.
\end{proof}
\end{thm}

 By  duality, the condition that a simple curve on a closed surface $\Sigma$ represents zero in $H_1(\Sigma;\ZZ)$ is equivalent to the condition that the complement of the curve has two components. This criterium is also appropriate to treat the case of a punctured surface $\Sigma$: one applies it to the closed surface $\Sigma'$ obtained from
 $\Sigma$ by either capping all its punctures by discs or shrinking boundary circles to points.

Let $P$ be a pair of pants and $i:P \hookrightarrow \Sigma$ be an inclusion into an oriented surface.

\begin{lemma}
$P$ is fashionable if and only if 
the map $i_*:H_1(P;\ZZ) \to H_1(\Sigma;\ZZ)$ is injective.
\begin{proof}
It is clear that if $i_*$ is injective, then $P$ is fashionable. We need to prove the opposite 
implication. Assume that $P$ is fashionable. Consider the complement $P^c := \Sigma \setminus P $. 
Since $P$ is fashionable, $P^c$ is connected. There is a decomposition $\Sigma=\bar P \cup {P^c}$.
Let $z\in \ker(i_*)$. 
From the Mayer--Vietoris long exact sequence of the decomposition, it follows that
$P$ has two boundary circles such that $z=a_1[C_1]+a_2[C_2]\in H_1(\bar P;\ZZ)$, and such that 
$a_1[C_1]+a_1[C_2]=0\in H_1({P^c};\ZZ)$. These homology classes correspond to the homology classes
of two punctures in $H_1(P^c;\ZZ)$. 
But since $P^c$ is connected, the only relation satisfied by the homology classes of the  
punctures of $P^c$ is that the sum of all of them is trivial. Since $P^c$ has at least three punctures,
we obtain that $a_1=a_2=0$.
\end{proof}
\end{lemma}

One could ask if there exist fashionable pair-of-pants decompositions on a surface $\Sigma$. Actually, they are quite common. At least for closed $\Sigma$, there is always a finite process that makes any given pair-of-pants decomposition fashionable. Let us describe an elementary operation that is useful for this purpose. Suppose that a given pair of pants $P\subset \Sigma$ in a
decomposition of $\Sigma$ is unfashionable because (at least) one of its interior boundary circles $S \subset \partial P \subset \Sigma$  is trivial in $H_1(\Sigma;\ZZ)$; a necessary condition 
is that this particular circle also occurs as boundary of another pair of pants $P'$ of the decomposition. The union $T:=\bar P \cup_S \bar P'  \subset \Sigma$ is homeomorphic to a 2-sphere with four disc punctures --- or equivalently, to one of the T-shirts depicted in Fig.~1. A {\em T-shirt flip} on $S$ consists of two steps: first, for each of $P,P'$ we mark one of the two boundary circles disjoint from $S$ as a {\em sleeve}; then, we replace $S$ by a new embedded
circle $S'  \subset T$ which defines two new pairs of pants $ P^\wedge,P^\vee \subset T$ with $T=\bar P^\wedge \cup_{S'}  \bar P^\vee$, in such a way that  the sleeves remain separated by $S'$; see Fig.~1. One gets three types of such flips up to diffeomorphisms of $T$ relative to $\partial T$, depending on the choice of sleeves. 

\begin{figure}[tb] \label{figTshirt}
\vspace{20pt}
\includegraphics[width=14cm,angle=0]{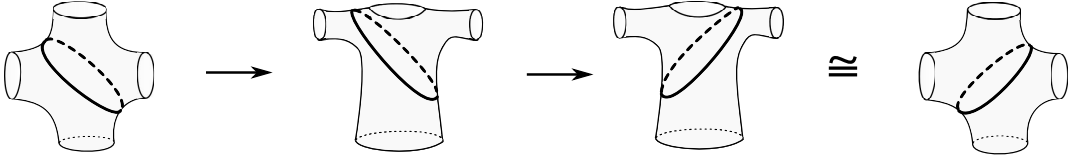} \\[-0pt]  \label{figPochhammer}
\caption{A T-shirt flip on the common boundary circle of two pairs of pants.}
\end{figure}

\begin{proposition} \label{fashion}
Let $\Sigma$ be an orientable surface of negative Euler characteristic. There is a
fashionable pair-of-pants decomposition of $\Sigma$.
\begin{proof}
If the Euler characteristic of $\Sigma$ is $-1$, $\Sigma$ is either a sphere with four holes or a torus 
with one hole, and any pair-of-pants decomposition will do.
We claim inductively that the theorem is true for surfaces of Euler characteristic greater or equal to $-d$. 
Let $\Sigma$ be an orientable surface of Euler characteristic $-d$. Pick a pair-of-pants decomposition. If
every boundary curve in this pair-of-pants decomposition is non-separating, we are done.
So suppose that we can find a simple curve $S$ that separates $\Sigma$ into two surfaces $\Sigma_1$ and $\Sigma_2$ of Euler characteristic each at least $-d+1$. By induction, we can find a fashionable pair-of-pants decomposition on each $\Sigma_i$. These decompositions combine to a pair-of-pants decomposition with one single separating boundary curve $S$. The two pair of pants adjacent to $S$ form a 2-sphere with four holes. Flip the pair-of-pants decomposition of this 2-sphere (here, any T-shirt flip of $S$ will do). Now all of the boundary curves are non-separating.
\end{proof}
\end{proposition}

For closed surfaces, the complement of the pairs of pants in a pair-of-pants decomposition (i.e.\ the collection of all boundary circles) is referred to as a marking. Hatcher and Thurston showed in the Appendix of ~\cite{HT} that markings are always related to each other by a finite number of moves of four types (I to IV), up to isotopy. Moves I and II are produced
 by our T-shirt flips, whereas move III can be performed by composing T-shirt flips with move IV. Combining their result with Proposition~\ref{fashion}, we conclude that
any given pair-of-pants decomposition of a closed surface can be rendered fashionable by a finite concatenation of T-shirt flips and Hatcher--Thurston type-IV moves.

In the light of Theorem~\ref{whyfashion}, it is natural to ask whether a concrete description of  generators for the twisted 1-homology groups of fashionable pairs of pants can be given. This is provided by our next result.

\begin{thm}[Pochhammer principle] \label{Pochhprinc}
 Let $P\subset \Sigma$ be a fashionable pair of pants.  
Let $f:S^1\to P$ be a curve homotopic to the commutator of two of the boundary circles.
Then $f_*:\HH_1(S^1)\to \HH_1(\Sigma)$ is an isomorphism.
\begin{proof}
  Since a pair-of-pants is homotopy equivalent to a wedge $S^1\vee S^1$,
  this follows from Remark~\ref{homotopy.lemniscates}.
\end{proof}
\end{thm}

Our discussion motivates the introduction of the following concept\footnote{On the twice-punctured complex plane $\mathbb{C} \setminus \{ 0,1\}$, homeomorphic to a pair of pants, the curve depicted on the lower left-hand side of Fig.~2 was used by Ludwig Pochhammer in his study of the Euler ${\rm B}$-function~\cite[p.~507]{Poc}. In complex analysis, there is some tradition in referring to this type of curves as Pochhammer contours, even though it has been recognized that they appeared in Camille Jordan's {\em Cours d'Analyse}~\cite{Jor} prior to Pochhammer's work (cf.~\cite[p.~256]{WhiWat}).}.

\begin{defn} \label{defPochh}
Let $P$ be a pair of pants in an oriented surface $\Sigma$. A {\em Pochhammer curve} in $P$ is a loop $\lambda: S^1 \rightarrow P$ such that $\lambda$ determines the trivial class in $H_1(P;\ZZ)$, but the induced map $\lambda_*:\HH_1(S^1) \rightarrow \HH_1(\Sigma)$ on $\alpha$-twisted homology of the Hurewicz cover of $\Sigma$ is injective.
\end{defn}

\begin{figure}[hb] \label{figPochhammer}
\vspace{30pt}
\includegraphics[width=12cm,angle=0]{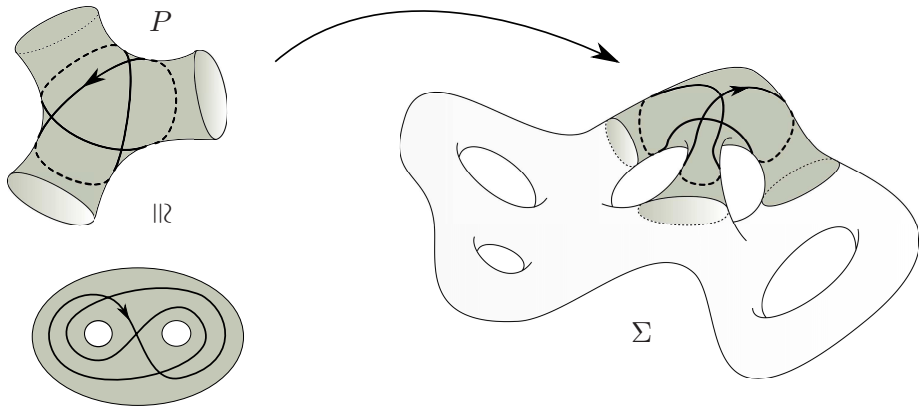}   \label{figPochhammer}\\
\vspace{-155pt}\hspace{-230pt}$P$\\
\vspace{55pt}\hspace{-230pt}$\vertiso$\\
 \vspace{30pt} \hspace{130pt}$\Sigma$\\
\vspace{30pt}
\caption{A Pochhammer curve in a fashionable pair of pants $P\hookrightarrow \Sigma$.}
\end{figure}

Even though this definition has been made regardless of $P$ being fashionable or not, our primary interest is in the case where $P$ is part of a fashionable pair-of-pants decomposition of $\Sigma$. A concrete example of a Pochhammer curve in a fashionably embedded pair of pants $P\hookrightarrow \Sigma$ is depicted in Fig~2. In the
model represented on the upper left-hand side, isotopic to the classical planar model below, the curve is approximately a geodesic for the metric induced by the embedding $P\hookrightarrow \RR^3$ in Euclidean space suggested by the drawing.

One could wonder what happens if a pair-of-pants decomposition of $\Sigma$ fails to be fashionable. 
Suppose that two pairs of pants $P_1,P_2$ in the decomposition have a common boundary which is a homologically trivial simple curve $S=\bar P_1\cap \bar P_2$; see Fig.~3 for a concrete example. Both inclusions
$i_1:S\hookrightarrow  P_1$ and $i_2:S\hookrightarrow  P_2$ induce isomorphisms on $\HH_1$.   
It follows that the images of the maps $\HH_1(P_1) \to \HH_1(\Sigma)$ and
$\HH_1(P_2) \to \HH_1(\Sigma)$ agree, so we cannot have that $\HH_1(\Sigma)$ decomposes as a direct sum of the subspaces given as images of the twisted homology of the individual pairs of pants $P_1,P_2$. 

\begin{figure}[tb] \label{figPoPs}
\vspace{20pt}
\includegraphics[width=8cm,angle=0]{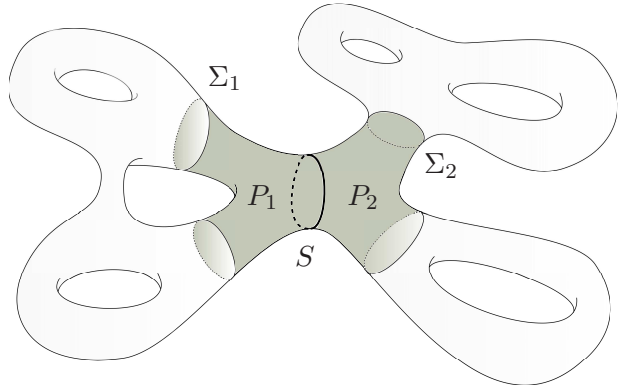} \\[-0pt]  \label{figPochhammer}
\vspace{-126pt}\hspace{-66pt}$\Sigma_1$ \\
\vspace{20pt}\hspace{96pt}$\Sigma_2$\\
\vspace{-1pt}
\hspace{0pt}$P_1$\hspace{26pt}$P_2$\\
\vspace{10pt}\hspace{-6pt}
$S$
\vspace{48pt}
\caption{Two unfashionable pairs of pants $P_1, P_2 \subset \Sigma:=\Sigma_1\cup_S \Sigma_2$.}
\end{figure}

Suppose also that we are given a homologically trivial simple curve
$S\subset \Sigma$. The curve will cut $\Sigma$ into two surfaces $\Sigma_1,\Sigma_2$ with common boundary $S$; see again Fig.~3 for an illustration. Let us assume that both $\chi(\Sigma_j)$ are negative. Then, by Proposition~\ref{fashion}, each $\Sigma_j$ can be given a
pair-of-pants decomposition such that all boundary curves in the decomposition, except for $S$, define nontrivial homology classes in $\Sigma$. In general, the image of $\HH_1(S)$ in $\HH_1(\Sigma)$ will be the sum of the classes defined by the two adjacent pairs of pants $P_1, P_2$ sharing $S$ as boundary circle. In the example depicted in Fig.~3, the ranks of the maps $H_1(P_j;\ZZ)\rightarrow H_1(\Sigma;\ZZ)$ are 1 for $j=1$ and 0 for $j=2$, so both $P_1$ and $P_2$ will contribute to this sum.

\section{$L^2$-Betti numbers and $\alpha$-twisted homology} \label{secL2Bettis}

As we mentioned in the introduction to this paper, there are essentially three versions of the $L^2$-Betti numbers of a covering space with infinite discrete group of deck transformations. In this section, we will provide a short account of the analytic and algebraic definitions, and review an alternative description of the algebraic viewpoint for the case of Abelian covers, which
connects to the definition we have given of twisted homology of $\alpha$-covers, without claiming originality. Among the references where the reader may retrieve more detail to fill out gaps in our
necessarily brief exposition are \cite{Lin,Rei,Lue}.

The analytic definition of the $L^2$-Betti numbers is the one directly relevant to the applications in mathematical physics that we are primarily interested in, and which we shall introduce
the reader to in section~\ref{physics} below. Suppose $M$ is a Riemannian manifold, and that there is a normal subgroup
$N$ of its fundamental group with discrete infinite quotient $\Gamma:=\pi_1(M)/N$; then we consider the corresponding Galois cover $p: \tilde M \rightarrow M$ with the pulled-back Riemannian structure and the isometric $\Gamma$-action via deck transformations. The metric on $\tilde M$ defines $L^2$-inner products on all spaces of tensors, and in
particular on the space of differential $i$-forms $\Omega^i(\tilde M)$. Let 
$$\mathcal{H}_{(2)}^i(\tilde M):= \{ \omega \in \Omega^i(\tilde M): \Delta \omega= 0, \| \omega \|_{L^2} <\infty \},$$
where $\Delta= {\rm d} {\rm d}^* + {\rm d}^* {\rm d}$ is the Laplace operator associated to the pulled-back metric (which is used to define the adjoint operators).  If we denote by
$L^2\Omega^i_c(\tilde M)$ the completion of the subspace of compactly supported $i$-forms on $\tilde M$ with respect to the $L^2$-inner product, there is an isometric
inclusion $\mathcal{H}_{(2)}^i(\tilde M) \hookrightarrow L^2\Omega^i_{c}(\tilde M)$ for each $i$; a theorem of Dodziuk~\cite{Dod} establishes that this map induces an
isomorphism between $\mathcal{H}_{(2)}^i(\tilde M)$ and the de Rham $L^2$-cohomology group
$$
H_{(2)}^i(\tilde M):= {\rm ker} ({\rm d}: \Omega^i_c(\tilde M)\rightarrow L^2\Omega^{i+1}_c(\tilde M))/
\overline{{\rm im} ({\rm d}: \Omega^{i-1}_c(\tilde M)\rightarrow L^2\Omega^{i}_c(\tilde M))},
$$
where the overline denotes completion with respect to the $L^2$-inner product. 
Now $\mathcal{H}_{(2)}^i(\tilde M)$ is a Hilbert module over the group von Neumann algebra $\mathcal{N}(\Gamma):=\mathcal{B}(\ell(\Gamma)^\Gamma)$, 
whose elements are bounded operators on the Hilbert space of square-integrable sequences $\ell^2(\Gamma)$ which are equivariant for the natural $\Gamma$-action.
There is a unique dimension function ${\rm dim}_{\mathcal{N}}(\Gamma)$ for such modules, taking values in $[0,\infty]$, called the Murray--von Neumann dimension~\cite{MurvNeu,Lue},
which extends the obvious trace in $\mathcal{B}(\ell^2(\Gamma))$ and satisfies natural properties. So one can set
\begin{equation}\label{analyticdef}
b^i_{(2)}(\tilde M,\Gamma):=\dim_{\mathcal{N}(\Gamma)} \mathcal{H}_{(2)}^i(\tilde M)
\end{equation}
and obtain sensible invariants of $\tilde M$ as  a $\Gamma$-space, which turn out to be homotopy invariant; these are the analytic $L^2$-Betti numbers. Atiyah's original definition~\cite{AtiEO} 
may seem more intuitive, since it does not explicitly use $\mathcal{N}(\Gamma)$: it expresses the quantity above as the limit of an integral over a fundamental domain
$F$ for the $\Gamma$-action on $\tilde M$,
\begin{equation}\label{Atiyahsdef}
b^i_{(2)}(\tilde M,\Gamma)=\lim_{t\rightarrow \infty} \int_F {\rm tr}_{\Lambda^i {\rm T}^*_x \tilde M}\left({\rm e}^{-t \Delta}\right) \, {\rm d}x.
\end{equation}
The integrand is the trace of the quadratic form associated  to the heat kernel ${\rm e}^{-t \Delta}$ on $i$-forms, restricted to the diagonal. In the  limit $t\rightarrow \infty$, this operator can be intuitively thought of  as a pointwise projection onto the space of harmonic $k$-forms, but in (\ref{Atiyahsdef}) the projection over global differential $i$-forms is being regularized to a `density of projection' with respect to the $\Gamma$-action as one restricts the integration to the fundamental domain.

Whereas the analytic version (\ref{analyticdef})--(\ref{Atiyahsdef}) of the $L^2$-Betti numbers provides perhaps the most direct geometric insight, it is arguably the most unwieldy, and the best way to calculate the invariants may
be to reinterpret them in terms of either of the other two definitions before embarking in actual computations. This strategy is always possible if the space $M$  in question is
a compact manifold, for then all the three definitions are equivalent~\cite{Lue}.  Since the algebraic definition does not depend on any (e.g.~Riemannian) extra structure, this also
shows that the analytic $L^2$-Betti numbers are independent of the metric structure if $M$ is compact. The standard definition of algebraic $L^2$-Betti numbers for a cover $\tilde M $ 
of $M$ with group of deck transformations $\Gamma$ uses again the Murray--von Neumann dimension:
\begin{equation}\label{algebraicdef}
b_i^{(2)}(\tilde M, \Gamma):= \dim_{\mathcal{N}(\Gamma)} H^\Gamma_i(\tilde M;\mathcal{N}(\Gamma)).
\end{equation}
Here, $H^\Gamma_i(\tilde M;\mathcal{N}(\Gamma))$ denotes the $i$-th $\Gamma$-equivariant  homology group with values in the von Neumann algebra
of $\Gamma$, which is understood as the homology of the ${\mathcal{N}}(\Gamma)$-chain complex
$$
C^{\rm sing}_*(\tilde M)\otimes_{\ZZ [\Gamma]}  \mathcal{N}(\Gamma),
$$
where $C^{\rm sing}_*(\tilde M)$ denotes the singular chain complex of right $\ZZ[\Gamma]$-modules.

There is a body of results relating the algebraic version  (\ref{algebraicdef}) of the $L^2$-Betti numbers in certain situations to dimensions over ring extensions of $\CC [\Gamma]$ alternative
to $\mathcal{N}(\Gamma)$. The basic signpost~\cite[p.~1]{Rei} is the following square diagram of ring extensions:
\[
\begin{CD}
\CC[\Gamma]  @>>>  \mathcal{N}(\Gamma)\\
@VVV @VVV \\
\mathcal{D} (\Gamma) @>>> \mathcal{U}(\Gamma)\\
\end{CD}
\]
In the lower row, $\mathcal{U}(\Gamma)$ is the algebra of affiliated operators of $\Gamma$, defined as the Ore localization~\cite{Ore, Rei} of $\mathcal{N}(\Gamma)$ with respect to the
multiplicative subset of non-zero divisors, whereas $\mathcal{D}(\Gamma)$ stands for the smallest division-closed intermediate ring between $\CC[\Gamma]$ and $\mathcal{U}(\Gamma)$. In \cite[Thm.~7]{Lin}, Linnell studies the ring $\mathcal{D} (\Gamma)$ for a large class of groups $\Gamma$. In particular, this class contains the free Abelian groups 
$\Gamma = A := \ZZ^n$ which play a central role in our paper. For this particular set of examples, the corners of the diagram above become familiar commutative $\CC$-algebras:
\[
\begin{CD}
\CC[z_1^\pm,\ldots, z_n^{\pm}] @>>> L^{\infty} (\TT^n) \\
@VVV @VVV \\
 \CC(z_1,\ldots, z_n) @>>> L(\TT^n)\\
\end{CD}
\]
Here, $\TT^n:=(S^1)^n$ is the $n$-torus equipped with its Haar measure, whereas $L(\TT^n)$ denotes measurable complex-valued functions, identified whenever they agree almost everywhere.
The transition between the two diagrams is to be understood via Fourier transform, which provides an identification of complex Hilbert spaces $\ell^2(\ZZ^n) \cong L^2(\TT^n)$. Note that $L^2(\TT^n)$ is a module over the ring $L^\infty(\TT^n)$ (acting by multiplication), which models the von Neumann algebra $\mathcal{N}(\ZZ^n)$. We use the embedding $S^1\subset \CC$ with a standard complex coordinate to describe the group algebra $\CC[\ZZ^n]$ as a space of Laurent polynomials.
In this language, the division closure $\mathcal{D}(\ZZ^n)$ becomes the ordinary quotient field of rational functions $\CC(z_1,\ldots, z_n)$ restricted to the real torus $\TT^n\subset (\CC^\times)^n$, acting on the Hilbert $L^\infty(\TT^n)$-module $L^2(\TT^n)$ once again by multiplication. We shall abbreviate, according to our previous notation,
 $$\CC(z_1,\ldots,z_n) \cong \CC(A) = \FF_\CC.$$

The crucial result we want to highlight in this section is that in the case $\Gamma=A:=\ZZ^n$  for $n={\rm rk}\, H_1(M;\ZZ)$, and letting $\alpha$ stand for the Hurewicz map of $M$, one has:

\begin{proposition} \label{voodoo}
$b_i^{(2)}(\tilde M, A)=\dim_{\FF_\CC} \HH_i(M;\FF_\CC)= \dim_\FF \HH_k(M).$
\begin{proof}
This statement is a consequence of the general fact (which holds for arbitrary $\Gamma$)
\begin{equation} \label{Reich}
b_i^{(2)}(\tilde M, \Gamma)= \dim_{\mathcal{U}(\Gamma) } H^\Gamma_i(\tilde M,\mathcal{U}(\Gamma))
\end{equation}
obtained in Reich's PhD thesis~\cite[Prop.~4.2.(ii)]{Rei}, where $\dim_{\mathcal{U}(\Gamma) }$ is the natural extension of the Murray--von Neumann dimension function~\cite{MurvNeu}
to arbitrary $\mathcal{U}(\Gamma)$-modules introduced in~\cite[Prop.~3.2]{Rei}.

To see how this result implies the proposition in the case $\Gamma=A$, consider the singular chain complex $C^{\rm sing}_*(\tilde M)$ of $\CC[A]$-modules and extend its coefficients to the division closure $\mathcal{D}(A) \cong \CC(A)$ which was described above. Taking the homology of the resulting complex, one obtains a $\CC(A)$-vector
space at each degree $i$, which we shall assume finite (for simplicity); let us denote by $\beta_i$ the corresponding dimension over $\CC(A)$:
$$
H_i(C^{\rm sing}_*(\tilde M) \otimes_{\CC[A]} \CC(A)) \cong \CC(A)^{\beta_i}.
$$
Now we tensor the complex once again to obtain
$$
\left( C^{\rm sing}_*(\tilde M)\otimes_{\CC[A]} \CC(A)\right) \otimes_{\CC(A)} \mathcal{U}(A) \cong C^{\rm sing}_*(\tilde M)\otimes_{\CC[A]} \mathcal{U}(A).
$$
Since tensoring over a field is an exact functor, we obtain homology groups
$$
H_i(C^{\rm sing}_*(\tilde M) \otimes_{\CC[A]} \mathcal{U}(A)) \cong \mathcal{U}(A)^{\beta_i}.
$$
The basic normalization property $\dim_{\mathcal{U}(A)} \mathcal{U}(A) = 1$ together with (\ref{Reich}) imply that $b_i^{(2)}(\tilde M, \Gamma)=\beta_i$. But by construction
 $\beta_i$ is also equal to $\dim_{\FF_\CC} \HH_i(M;\FF_\CC)=\dim_\FF \HH_1(M)$.
\end{proof}
\end{proposition}

We now want to revisit Atiyah's key example (or rather its Abelian version) mentioned in the introduction. Since the Abelian cover $\tilde \Sigma$ of a compact Riemann
surface is a co-compact space for the action of $\Gamma=H_1(\Sigma;\ZZ)$, by the discussion above its analytic and algebraic $L^2$-Betti numbers agree, and by Proposition~\ref{voodoo} they also agree with the $\FF$-dimension of the $\alpha$-twisted homology for the Hurewicz map of the surface:
$$
b^i_{(2)}(\tilde \Sigma,\Gamma)=b_i^{(2)}(\tilde \Sigma,\Gamma)=\dim_\FF \HH_i(\Sigma).
$$
In particular, since our discussion in section~\ref{seclemniscates} provided explicit $\FF$-bases for $\HH_1(\Sigma)$ in terms of Pochhammer curves, we have now obtained the intuitive interpretation of Atiyah's result (\ref{Atiyah}) advertized in the introduction. We emphasize that using this type of argument one calculates the nontrivial invariants, in this case $b^1_{(2)}(\tilde \Sigma,\Gamma)$, without having to rely on vanishing results for the other $L^2$-Betti numbers.

\section{Charged particles on closed surfaces} \label{physics}

This final section illustrates how the topological considerations of the previous sections can be applied to  two-dimensional gauge theory.

\subsection{Aharonov--Bohm effect for supersymmetric particles on surfaces} \mbox{}\\[-5pt] \label{secAB}

Let us recall the setup of $N=(2,2)$ supersymmetric quantum mechanics~\cite{WitSMT,HV} on a compact orientable surface $\Sigma$ of genus $g>1$ equipped with a K\"ahler metric $g_\Sigma$.
Classically, this system is described by pairs $(\phi,\psi)$ consisting of `bosonic' paths $\phi: \RR \rightarrow \Sigma$ together with their `fermionic' complex deformations $\psi \in \Gamma(\RR,\phi^*{\rm T}\Sigma\otimes \CC)$. The dynamics of these paths is dictated by a variational principle for the Lagrangian $L:{\rm T}C^\infty(\RR,\Sigma)\otimes\CC  \rightarrow \RR$ given by
$$
L[(\phi,\psi)]:=\frac{1}{2}\, g_\Sigma(\dot \phi,\dot \phi)+ \frac{\rm i}{2}\, \omega_{\Sigma}\left(\bar \psi, \nabla_{\dot\phi} \psi\right)+ g_\Sigma\left(\phi_*\bar \psi,R_\nabla(\phi_* \psi,\phi_*\bar \psi)\phi_*\psi\right) ,
$$
where $\nabla$ is the Levi--Civita connection of $g_\Sigma$, $\omega_\Sigma$ the associated K\"ahler form and $R_\nabla$ the Riemann curvature tensor of $\nabla$; we are extending $\CC$-linearly  all covariant tensors to ${\rm T}\Sigma\otimes \CC$.
This is the simplest example of a supersymmetric sigma-model (with source $\RR$ and target $\Sigma$). Its Euler--Lagrange equations are the `fuzzy' geodesic equations
$$
\nabla_{\dot \phi} \dot \phi =R_\nabla(\phi_*\psi,\phi_*\bar\psi)\dot \phi,\qquad \nabla_{\dot \phi} \psi=0=\nabla_{\dot \phi} \bar \psi.
$$

The canonical quantization of this system produces the infinite-dimensional quantum Hilbert space of complex-valued forms $\mathcal{H}=\Omega^*(\Sigma;\CC)$, with inner product induced by $g_\Sigma$ and the usual inner product on $\CC$. Quantum states of a supersymmetric particle are represented by vectors in $\mathcal{H}$, referred to as {\em waveforms}, and the supersymmetric viewpoint uses the splitting $\mathcal{H}=\mathcal{H}^+\oplus \mathcal{H}^-$ into forms of even/odd degree which correspond to bosons/fermions, respectively.
This Hilbert space provides a super-representation of the 8-dimensional $N=(2,2)$ supersymmetry Lie super-algebra generated by the Laplacian $\Delta={\rm d}{\rm d}^*+{\rm d}^*{\rm d}$, the Dolbealut operators $\partial, \bar\partial$ and their adjoints, as well as the Lefschetz $\mathfrak{sl}_2\CC$-triple generated by wedging with $\omega_\Sigma$, its adjoint and their commutator~\cite{Huy}. The operator $\Delta$ plays the role of quantum Hamiltonian, and the spectral decomposition $\mathcal{H}=\bigoplus_{E\in {\rm Spec}(\Delta)} \mathcal{H}_E$ has special significance. A consequence of $\Delta=({\rm d}+\rm{d}^*)^2$ is that one has ${\rm Spec}(\Delta)\subset [0,\infty[$ (this is the spectrum of  {\em energies} of the model);
moreover, whenever $E>0$, ${\rm d}+{\rm d}^*$ restricts to an isomorphism $\mathcal{H}_E^+\cong \mathcal{H}_E^-$ where $\mathcal{H}_E^\pm := \mathcal{H}_E\cap \mathcal{H}^{\pm}$.  The eigenspace $\mathcal{H}_0$ is the
subspace of harmonic forms, and it is referred to as the sector of {\em ground states} or vacua (i.e. quantum states of minimal energy). We can model the elements of $\mathcal H_0$
as de Rham cohomology classes via standard Hodge theory. Note that the two summands $\mathcal{H}_0^\pm$ are not necessarily isomorphic, and the formal difference between
them represents a net balance of unpaired fermions over the whole $\mathcal{H}$. The dimension of this formal difference is 
${\rm tr}_{\mathcal{H}}(-1)^{\rm deg}=\chi(\Sigma)=2-2g<0$.

Now we want to twist the whole picture by introducing  ${\rm U}(1)$-connections ${\mathcal A}$  in bundles over $\Sigma$. For now, we can think of  $\mathcal A$ as a background 
gauge field to which the supersymmetric particle may couple if it has a charge.
The simplest possibility is to demand that ${\mathcal A}$ be flat, and in this case the interaction occurs non-locally via holonomies along curves in $\Sigma$ (so we can think of the
field produced by ${\mathcal A}$ as being undetectable locally, but nontrivially threaded over the `holes' of $\Sigma$). This is precisely the setup for the Aharonov--Bohm effect, which is also used to model statistical phases in two-dimensional systems. To implement this, one constructs  a rank-one local system over $\Sigma$ by specifying 
a character $\xi$ in
\begin{equation} \label{repvariety}
 \mathcal{R}:= {\rm Hom}(\pi_1(\Sigma),{\rm U}(1))\cong{\rm U}(1)^{2g}=\TT^{2g}.
\end{equation}
Concretely, we assign to $\xi$ the Hermitian line bundle $\mathcal{E}_\xi:=\tilde\Sigma\times_\xi\CC$ with its $\xi$-twisted flat connection, where $\tilde \Sigma$ is the universal cover of $\Sigma$ and $\CC$ carries the usual Hermitian inner product. Since the structure group ${\rm U}(1)$ is Abelian, the holonomy  along a loop will only depend on the homology class of that loop. This means that each local system $\mathcal{E}_\xi$ will become trivial already on the maximally connected Abelian cover of $\Sigma$, which coincides with the Hurewicz cover defined in section~\ref{secalpha}. Hence we can, and will, replace the universal cover in the argument above by this Hurewicz cover, which will be denoted $p_\alpha: \tilde \Sigma\rightarrow \Sigma$ from now on. The quantum Hilbert space $\mathcal{H}$ that we considered above should now be replaced by the twisted version $\mathcal{H}^{(\xi)}:=\Omega^*(\Sigma;\mathcal{E}_\xi)$, with the untwisted (or uncharged) situation being recovered when $\xi$ is the trivial representation.

Physically, the holonomies have measurable effects on the charged particle which can be detected in the form of interference patterns; but such measurements are never totally accurate, so it is more satisfactory to work with values of $\xi$ that are smeared within a (possibly very small) subset in $\mathcal R$, rather than restricted to a precise value.
A natural further step is to promote the character $\xi$ to be an internal quantum number of the system, and consider a quantum master space 
$\prod_{\xi \in \mathcal{R}}\mathcal{H}^{(\xi)}$ incorporating all characters; the quantum operator corresponding to $\xi$ is usually referred to as a Wilson loop. Unfortunately, such a
master space would not be a Hilbert space. To remedy this, one needs to regularize by working  with `wave-packets' weighted over the representation variety $\mathcal R$. 
This will be possible because $\mathcal{R}=\TT^{2g}$ carries a natural measure --- in this case, it is a compact Lie group and the Haar measure can be used for this purpose. 

To make this idea more concrete, we lift the twisted waveforms in each $\mathcal{H}^{(\xi)}$ to the Hurewicz cover $\tilde \Sigma$. Note that $\tilde \Sigma$ has an
action of $\ZZ^{2g}$, making the Hilbert space $L^2 \Omega_c^*(\tilde \Sigma;\CC)$ (where the $L^2$-inner product is with respect to the pulled-back metric $p_\alpha^*g_\Sigma$) into a module over the corresponding group von Neumann algebra $\mathcal{N}(\ZZ^{2g})\cong L^\infty(\TT^{2g})$. 
Given any fundamental domain $F\subset \tilde\Sigma$ of the $\Gamma$-action, the decomposition~\cite{AtiEO}
$$
L^2 \Omega_c^*(\tilde \Sigma;\CC)\cong \Omega^*(F;\CC)\otimes \ell^2(\Gamma) \cong \Omega^*(\Sigma;\CC)\otimes L^2(\TT^{2g})
$$
substantiates the interpretation of $L^2$-integrable forms on $\tilde \Sigma$ as wave-packets of forms on $\Sigma$ valued in line bundles, for a spectrum of twistings parametrized by $\mathcal R$. For example, to
obtain wave-packets that are supported on measurable subsets $M\subset  \TT^{2g}$ one can project using the charactetistic functions $\chi_M $,
which are  idempotents in $\mathcal{N}(\TT^{2g})$.
Working with weights supported on a small subset $M \ni  \xi $, one gets wave-packets of waveforms in each degree $i$ that are localized around $\xi$.
In particular, we obtain wave-packets of ground states from $L^2$-normalisable harmonic forms on the Hurwitz cover. Since the space $\tilde \Sigma$ is no longer compact, 
the appropriate version of Hodge theory in this setting relates  $\mathcal{N}(\Gamma)$-modules of harmonic forms to $L^2$-cohomology, as already stated in section~\ref{secL2Bettis}. Whenever we localize with respect to a supporting subset $M$ with nonzero measure, we are treating an infinite number of characters at a time, and so these modules will be infinitely generated $\CC$-vector spaces. Now we can extract information from them in the form of their $L^2$-Betti numbers with respect to the $\Gamma$-action on $\tilde \Sigma$. Since in this case $\Gamma$ is an Abelian group without torsion, after dividing by ${\rm Vol}(M)$ one always ends up with integer quantities $b^i_{(2)}(\Sigma,\Gamma)$ that have a probabilistic interpretation of renormalized dimensions of spaces of harmonic $L^2$-waveforms of degree $i$, for characters averaged over $M$. A more detailed account of these ideas and techniques will be given elsewhere~\cite{BokRomWeg}.
 
We can connect this viewpoint on charged supersymmetric particles with the topological discussion in this paper. 
We have shown in section~\ref{secL2Bettis} that we can compute the $L^2$-Betti numbers for a compact surface of genus
$g>1$ as the actual dimensions (over $\FF$) of $\alpha$-twisted homology groups $\HH_1(\Sigma)$ with respect to the Hurewicz map of $\Sigma$. More precisely, we are
now able to assign to any fashionable pair-of-pants decomposition $\{P_j\}_{j=1}^{2g-2}$ of $\Sigma$, decorated by arbitrary Pochhammer loops $\lambda_j:S^1\rightarrow P_j\subset \Sigma$ on each of its pairs-of-pants, a set of $b^1_{(2)}(\Sigma,\Gamma)=2g-2$ {\em Pochhammer vectors}
\begin{equation} \label{lemniskets}
|\lambda_j \!\varpropto\! P_j \rangle := \lambda_{j*} ([\widetilde{S^1}]) \in \HH_1(\Sigma; \CC),\qquad j=1,2,\ldots, 2g-2.
\end{equation} 
These vectors provide a useful device to represent states of charged quantum supersymmetric particles on $\Sigma$ by virtue of the following
 {\em sewing property}:

\begin{proposition} \label{propsewing}
If  $J\subset \{1,2,\ldots, 2g-2\}$, then
\begin{equation} \label{sewing} 
{\rm span}_{\FF_\CC} \left\{  |\lambda_j \!\varpropto\! P_j \rangle  \right\}_{j\in J} \cong \HH_1\left(\cup_{j\in J}\bar P_j; \CC \right) \cong \FF_\CC^{|J|}.
\end{equation}
In particular, the  set $\{  |\lambda_j \!\varpropto\! P_j \rangle  \}_{j=1}^{2g-2}$ provides an $\FF_\CC$-basis for the space $\HH_1(\Sigma;\FF_\CC)$. 
\begin{proof}
This follows from Theorem~\ref{decomposition} and the Pochhammer principle in Theorem~\ref{Pochhprinc}.
\end{proof}
\end{proposition}

Our discussion so far has been concerned with quantum one-particle states of the charged supersymmetric particle. For  $k$ identical  particles, the standard recipe in nonrelativistic quantum mechanics consists of prescribing first the statistics of the particles (bosonic, fermionic) and then constructing their quantum Hilbert space as the $k$-th exterior or symmetric power of the one-particle states, respectively, projecting out from the tensor product. In a supersymmetric system, the natural procedure is to treat waveforms of all degrees at the same time and
consider the whole tensor product of one-particle states. One alternative to this orthodoxy is to use configuration spaces, and this allows for the treatment of anyonic particles  (this
viewpoint has some tradition in the physics of two-dimensional systems~\cite{Wil}). In the discussion above, this would involve replacing $\Sigma$ by ${\rm Conf}_k(\Sigma):=(\Sigma^k\setminus \Sigma_{\Delta})/\mathfrak{S}_k$, where $\Sigma_\Delta$ is the diagonal and $\mathfrak{S}_k$ the symmetric group permuting the copies in the cartesian product. The noncompactness of these spaces, as well as the intricacy of their fundamental groups ${\mathcal B}_k(\Sigma)$ (surface braid groups in $k$ strands), causes difficulties; note also that this framework still requires the statistics of the multiparticles to be prescribed by hand.

\subsection{Vortices in gauged linear sigma-models}\mbox{}\\[-5pt] \label{vorticesGLSM}

We shall now argue that our discussion extends to the quantization of certain  $(1+2)$-dimensional gauge theories --- most immediately, to the study of their nonperturbative vacua. The theories that we have in mind are supersymmetric sigma-models with $\RR \times \Sigma$ (carrying the Lorentzian product metric ${\rm d}t^2-g_{\Sigma}$) as source, and target any K\"ahler manifold $(X,J,\omega_X)$ with a holomorphic Hamiltonian action of a compact  Lie group $G$. In this paper, we will be confining the discussion to linear gauged sigma-models and shall take $X=\CC^{r\times r}$  to be the $\CC$-vector space of $r\times r$ matrices with the usual Euclidean metric and action of $G={\rm U}(r)$ by left-multiplication.
At the bosonic level, the variables are a ${\rm U}(r)$-connection $A$ on a principal ${\rm U}(r)$-bundle $\mathcal P$ over $\Sigma$ and a section $u$ of the associated vector bundle $\mathcal{P} \times_{{\rm U}(r)}{\CC}^{r\times r}$.
The Lagrangians contain the standard sigma-model kinetic terms  associated to the $L^2$-norms of the time derivatives and time-components of the fields $(A,\phi)$ in the Lorentzian metric, and a  term of potential energy $V_{\tau,\lambda}$ which can be expressed as follows:
$$
V_{\tau,\lambda}[(A,u)] := \frac{1}{2} \int_\Sigma \left( |F_A|^2 + |{\rm d}_A\phi|^2  + \frac{\lambda}{4}|\mu_\tau \circ \phi|^2\right).
$$
The third summand is the Fayet--Iliopoulos term, which depends on two real parameters $\lambda$ and $\tau$. The latter is associated to the choice of moment map $$\mu_\tau(w):= -\frac{1}{2} \left(w \bar w^{\rm t}-\tau {\mathbf 1}_r\right)$$ for the ${\rm U}(r)$-action; note that  we are identifying $\mathfrak{u}(r)$ and its dual equivariantly. The identity
$$
V_{\tau,\lambda}[(A,u)]= 2 \pi \,{\rm deg}( \det u)  + \frac{1}{2}\int_\Sigma \left( |F_A+ (\mu\circ u)\,\omega_\Sigma|^2 + 2 |\bar\partial_A u|^2  + \frac{\lambda-1}{2} |\mu_\tau\circ u|^2 \right)
$$
identifies  the critical value $\lambda=1$ of the other parameter as the self-dual point. If the topological charge $ {\rm deg}( \det u)=:k$ is positive, one can express the minima of $V_{\tau, 1}$ by the system of first-order PDEs
\begin{equation} \label{vortexeqs}
\bar\partial_A u=0, \qquad \qquad F_A+ (\mu\circ u)\,\omega_\Sigma=0
\end{equation}
which are a particular case of the {\em vortex equations}. More than studying their solutions, it is important to describe the space
\begin{equation} \label{moduli}
{\mathcal M}^{\rm U(r)}_{\Sigma,k}:= \{ (A,u): \bar\partial_A u=0, \;  F_A+ (\mu\circ u)\,\omega_\Sigma=0, \; {\rm deg} \det \phi = k \} / {\rm Aut}_\Sigma (\mathcal{P})
\end{equation}
of gauge-equivalent solutions of a given charge $k>0$. In the situation where $\tau > \frac{4 \pi k}{{\rm Vol}(\Sigma)}$, the {\em moduli space} (\ref{moduli}) has been described in~\cite{BapNAV} and~\cite{BisRom}, for example.
It is a complex manifold carrying a nontrivial K\"ahler structure $\omega_{L^2}$, induced by the kinetic term of the bosonic sigma-model, which depends on $\Sigma, g_\Sigma, \tau, r$ and $k$. The corresponding K\"ahler metrics encode information about inter-vortex interactions: for instance, their geodesic flow approximates the classical dynamics for the bosonic sigma-models for $\lambda$ close to $1$ and small velocities~\cite{ManSut} --- see \cite{Stu} for the analysis in the case where $k=2$ and $\Sigma=\CC$.

It turns out that the bosonic sigma-models we have just described admit supersymmetric versions at the critical value $\lambda = 1$. There are two ways to twist~\cite{Wit-QFS} the Lagrangians with the fermions added to obtain functionals for globally defined fields on $\Sigma$, and they lead to topological field theories that were described in~\cite{BapTSM}. In this paper, we will be interested in the {\em A-twist} that makes use of the vectorial (global) circle $R$-symmetry~\cite{HV}. In the Lagrangian formulation of the corresponding TQFT, the path integrals of this model localize to the moduli spaces in (\ref{moduli}). This justifies the following strategy to study the A-twisted gauged sigma-models at low energies: one
replaces the supersymmetric $(1+2)$-dimensional field theories by a supersymmetric $1$-dimensional sigma-model  analogous to the one described in section~\ref{secAB}, the only difference being that the target is now taken to be the K\"ahler manifold ${\mathcal M}^{{\rm U}(r)}_{\Sigma,k}$ rather than $\Sigma$, and we end up by considering Hilbert spaces of waveforms on vortex moduli spaces.
In this context, the extension to twisted waveforms, incorporating holonomies in a moduli space of magnetically charged particles, is in the same spirit as
the generalization of magnetic monopoles in $3+1$ dimensions to dyonic particles that also possess electric charge~\cite{OliWes}.

In the Abelian case $r=1$, the space of 1-vortices is 
\begin{equation}\label{M1_r1}
{\mathcal M}^{{\rm U}(1)}_{\Sigma,1} \cong \Sigma
\end{equation}
as a complex manifold exactly like in section~\ref{secAB}, but now it is endowed with a nontrivial metric $g_{L^2}$ for each $\tau$, which is distinct from the metric $g_\Sigma$ used to define the
$(1+2)$-dimensional Lagrangian, or to write down
the vortex equations. The metric $g_{\Sigma}$ should be thought of
as the limit $\lim_{\tau \rightarrow \infty} g_{L^2}$ where the vortex becomes a point particle, whereas the other extreme $\tau \rightarrow \frac{4 \pi}{{\rm Vol}(\Sigma)}$ corresponds to the limit of `dissolved vortices' associated to the Bergman metric on the Riemann surface $\Sigma$, as discussed in~\cite{ManRom}. It was established in~\cite{BapNAV} that the generalization of (\ref{M1_r1}) for any $r$ subject to $\tau {\rm Vol}(\Sigma)> {4 \pi}$ is
$$
{\mathcal M}^{{\rm U}(r)}_{\Sigma,1} \cong \mathbb{P}^{r-1} \times \Sigma.
$$
Thus for the Abelian model, a 1-vortex is associated with a point on the surface (where the Higgs field $u$ has a simple zero), whereas for the non-Abelian case $r>1$, apart from
these `spatial moduli', there is an extra factor $\PP^{r-1}$ in the configuration space which parametrizes internal structures~\cite{BapNAV}. 

We can now extend the discussion of Aharonov--Bohm phases to the 1-vortex moduli spaces ${\mathcal M}^{{\rm U}(r)}_{\Sigma,1}$, with a view of
understanding ground states of charged  single particles in the quantum (1+2)-dimensional gauged linear sigma-models. 
Consider the $\alpha$-cover associated to the homomorphism
\begin{equation} \label{alphaproj}
\alpha: \pi_1 \left(\mathcal{M}^{{\rm U}(r)}_{\Sigma,1}\right) \rightarrow A:=H_1(\Sigma;\ZZ)
\end{equation}
obtained by composition of the isomorphism in $\pi_1$ induced by the projection $\PP^{r-1}\times \Sigma \rightarrow \Sigma$ with the Hurewicz map of $\Sigma$. 
Then we obtain the following result.

\begin{thm} \label{GLSM}
Let $\Sigma$ be a compact Riemann surface of genus $g>1$.  The $L^2$-Betti numbers of the moduli space of ${\rm U}(r)$-1-vortices 
for the $\alpha$-cover associated to (\ref{alphaproj}) are
\begin{equation}
b^i_{(2)}\left(\widetilde{\mathcal{M}}^{{\rm U}(r)}_{\Sigma,1}, H_1(\Sigma;\ZZ)\right)=
\left\{
\begin{array}{ll}
2g-2 & \text{ if $1\le i \le 2r-1$ and $i$ is odd,} \\
0 & \text{ otherwise}.
\end{array}
\right. 
\end{equation}
Given a pair-of-pants decomposition $\{P_j\}_{j=1}^{2g-2}$ decorated with Pochhammer curves $\lambda_j$, we can associate the following $\FF_\CC$-basis for each nontrivial
complex twisted homology group $\HH_i(\mathcal{M}^{{\rm U}(r)}_{\Sigma,1};\FF_\CC)$:
\begin{equation}\label{basisUr}
\psi_{i-1} \otimes \left|  \lambda_j \!\varpropto\! P_j \right\rangle\qquad i=1,\ldots 2r-1\text{ odd},\; j=1,\ldots, 2g-2,
\end{equation}
where $\psi_i$ are generators of the vector spaces $H_i(\PP^{r-1};\QQ )\cong \QQ$, and 
$ \left|  \lambda_j \!\varpropto\! P_j \right\rangle \in \HH_1(\Sigma;\FF_\CC)$ are the Pochhammer vectors defined in (\ref{lemniskets}).
\begin{proof}
The calculation of $L^2$-Betti numbers is a straightforward application of Theorem~\ref{thmKuenneth}, whereas the basis described  combines the
definition of Pochhammer vectors with the fact that the nontrivial homology groups of complex projective spaces are cyclic.
\end{proof}
\end{thm}

The basis  (\ref{basisUr}) displays a factorization of the ground states in the one-particle sector of the sigma model into spatial components associated to 
charged supersymmetric particles on $\Sigma$ (i.e.~the Pochhammer vectors in equation (\ref{lemniskets})), and internal bosonic waveforms associated to
closed complex submanifolds of the space of internal structures $\PP^{r-1}$.  This  type of factorization (obtained here from first principles by quantization of a classical moduli space) is  often assumed in  discussions of nontrivial vacua in quantum field theories. Note that all these extended states (which include nontrivial internal structures whenever $r>1$) are {\em fermionic}, just like the states of the charged supersymmetric point particle discussed in section~\ref{secAB}.

There are many directions in which to generalize the discussion in this section. Most immediately, one is  interested in understanding multiparticle states, for which $k>1$; they
can be obtained from quantizing moduli spaces of multivortices~\cite{BokRomWeg}. The simplest example is the Abelian case $r=1$, and one then deals with moduli spaces~\cite{Bra,GP}
$${\mathcal{M}^{{\rm U}(1)}_{\Sigma,k}}\cong {\rm Sym}^k(\Sigma):=\Sigma^k/\mathfrak{S}_k$$
with Abelian fundamental group isomorphic to $H_1(\Sigma;\ZZ)$. So in this case, once we think of charged particles, we are back to considering twisting line bundles
and smearing out waveforms using weights supported on the same character variety $\mathcal{R}$ as in (\ref{repvariety}).
The calculation of the $L^2$--Betti numbers~\cite{BokRomWeg} makes use of recent results on the topology of the universal cover  $\widetilde{{\rm Sym}}^k(\Sigma)$ obtained in~\cite{BokRomC}. The
basic result is
\begin{equation} \label{multip}
  b^i_{(2)}\left(\widetilde{\mathcal{M}}^{{\rm U}(1)}_{\Sigma,k} , H_1(\Sigma;\ZZ)\right) =  {2g-2 \choose k}  \delta^i_{k},
\end{equation}
where $\delta^i_k$ is the Kronecker delta, and this formula supports the independent interpretation of quantum vortices as fermionic particles given above. We should emphasize that the statistics of the quantum particles, in either of the two arguments,
naturally emerges from calculations derived from first principles, in contrast with the arbitrariness in prescribing  statistical properties for point particles in quantum mechanics.
A direct connection  between  formula~(\ref{multip}) and the traditional approach
to quantum multiparticle states of fermions will be provided in~\cite{RomWeg}. For the non-Abelian models, the picture is more intricate, as 
the spaces of internal structures coalesce nontrivially according to a fibration of the moduli space~\cite{BisRomNG}
$$
\mathcal{M}^{{\rm U}(r)}_{\Sigma,k} \rightarrow {\rm Sym}^k(\Sigma)
$$
with typical fibres depending on the natural filtration of ${\rm Sym}^k(\Sigma)$ by partitions of $k$.

More interesting structure emerges for gauged {\em nonlinear} sigma-models, where fundamental fermionic particles of different types coexist and can even merge nontrivially, giving rise to a much richer spectrum of 
quantum particles~\cite{RomWeg}. The fact that the moduli spaces are no longer compact in this situation leads to various difficulties, and their fundamental groups may also
become non-Abelian~\cite{BokRomDL}.
However, the description of the spatial quantum numbers provided by Pochhammer curves in pair-of-pants decompositions remains useful in this context. This language should provide 
helpful tools to study quantization of moduli spaces of gauge theories on Riemann surfaces more generally.

\vspace{15pt}
\noindent
{\sc Acknowledgements:} This work was inspired by activities developed by the second author at the Hausdorff Research Institute for Mathematics, University of Bonn,
in the framework of the JHTP `Mathematical Physics' in 2012. He would like to thank Thomas Schick for a useful discussion
and IMS at the National University of Singapore for hospitality during the programme `The Geometry, Topology and Physics of Moduli Spaces of Higgs Bundles'.

\bibliographystyle{numsty}

\begin{thebibliography}{9999}
                        \setlength{\itemsep}{0 pt}
                          \setlength{\parskip}{0pt}
                          \setlength{\parsep}{0pt}

\begin{small}


\bibitem{AtiEO}
\textsc{M.F. Atiyah}:
Elliptic operators, discrete groups and von Neumann algebras.  
{\sl Ast\'erisque} {\bf 32--33} (1976) 43--72
         
\bibitem{BapTSM}
\textsc{J.M. Baptista}: Twisting gauged non-linear sigma-models.
\newblock \textsl{JHEP} \textbf{0802} (2008) 096         
         
\bibitem{BapNAV}
\textsc{J.M. Baptista}: Non-abelian vortices on  compact Riemann surfaces.
\newblock \textsl{Commun.\ Math.\ Phys.} \textbf{291} (2009) 799--812

\bibitem{BisRom}
\textsc{I. Biswas, N.M.~Rom\~ao}:
Moduli of vortices and Grassmann manifolds.  
{\sl Commun. Math. Phys.} {\bf 320} (2013) 1--20       
                  
\bibitem{BisRomNG}
\textsc{I. Biswas, N.M.~Rom\~ao}:
A no-go theorem for nonabelionic statistics in gauged linear sigma-models;
(in preparation)        
         
\bibitem{BokRomC}
\textsc{M. B\"okstedt, N.M.~Rom\~ao}:
On the curvature of vortex moduli spaces.  
{\sl Math. Z.} {\bf 277} (2014) 549--573

\bibitem{BokRomDL}
\textsc{M. B\"okstedt, N.M.~Rom\~ao}:
Divisor links and fundamental groups of toric vortex moduli; (in preparation)

\bibitem{BokRomWeg}
\textsc{M. B\"okstedt, N.M.~Rom\~ao, C.~Wegner}:
$L^2$-invariants and supersymmetric quantum mechanics on vortex moduli spaces; (in preparation)

\bibitem{Bra}
\textsc{S.B. Bradlow}: Vortices in holomorphic line bundles over closed
K\"ahler manifolds. 
\newblock \textsl{Commun.\ Math.\ Phys.} \textbf{135} (1990) 1--17

\bibitem{Dod}
\textsc{J. Dodziuk}:
De Rham--Hodge theory for $L^2$-cohomology of infinite coverings.  
{\sl Topology} {\bf 16} (1977) 157--165

 
 \bibitem{FulAT}
\textsc{W. Fulton}: 
{\em Algebraic Topology: A First Course}, Springer-Verlag, 1995
    
 \bibitem{GP}
{\sc \'O. Garc\'\i a-Prada}: {A direct existence proof for the vortex
equations over a compact Riemann surface},
{\sl Bull. London Math. Soc.} \textbf{26} (1992) 88--96
   
         
\bibitem{Hat}
\textsc{A. Hatcher}: 
{\em Algebraic Topology}, Cambridge University Press, 2002

\bibitem{HT}
\textsc{A. Hatcher, W. Thurston},
   {\em A presentation for the mapping class group of a closed orientable
   surface},
   {\sl Topology} 
   {\bf 19},(1980), 221--237


\bibitem{HV}
\textsc{K. Hori, S. Katz, A. Klemm, R. Pandharipande, R. Thomas, C. Vafa, R. Vakil, E. Zaslow}:
{\it Mirror Symmetry}, American Mathematical Society, 
2003


\bibitem{Huy}
\textsc{D. Huybrechts}:
{\it Complex Geometry: An Introduction}, Springer-Verlag, 
2005

\bibitem{Jor}
\textsc{C. Jordan}:
{\it Cours d'Analyse de l'\'Ecole Polytechnique}, vol. III, Gauthier-Villars, 
1887


\bibitem{Lin}
\textsc{P.A. Linnell}:
Zero divisors and group von Neumann algebras.  {\sl Pacific J. Math.} {\bf 149} (1991) 349--363

\bibitem{Lue}
\textsc{W. L\"uck}: 
{\em $L^2$-Invariants: Theory and Applications to Geometry and K-Theory}, Springer-Verlag, 2002



\bibitem{ManRom}
\textsc{N.S. Manton, N.M.~Rom\~ao}:
Vortices and Jacobian varieties.  
{\sl J. Geom. Phys.} {\bf 61} (2011) 1135--1155


\bibitem{ManSut}
\textsc{N. Manton, P. Sutcliffe}: {\it Topological Solitons}, Cambridge University Press, 
2004


\bibitem{MurvNeu}
\textsc{F.J. Murray, J. von Neumann}: On rings of operators II.
\newblock \textsl{Trans. Amer. Math. Soc.} \textbf{41} (1937) 208--248

\bibitem{OliWes}
\textsc{D.I. Olive, P.C. West} (Eds.): {\it Duality and Supersymmetric Theories}, Cambridge University Press, 
1999


\bibitem{Ore}
\textsc{\O. Ore}:  Linear equations in non-commutative fields.
\newblock \textsl{Ann. Math.} \textbf{32} (1931) 463--477
\bibitem{Rei}
\textsc{H. Reich}: {\it Group von Neumann Algebras and Related Algebras}. PhD Thesis, University of G\"ottingen,
1998


\bibitem{Poc}
\textsc{L. Pochhammer}: Zur Theorie der Euler'schen Integrale.
\newblock \textsl{Math. Ann.} \textbf{35} (1890) 495--526


\bibitem{RomWeg}
\textsc{N.M.~Rom\~ao, C.~Wegner}:
$L^2$-Betti numbers and particle counting in a gauged nonlinear sigma-model; (in preparation)

\bibitem{Stu}
\textsc{D.M.A. Stuart}: Dynamics of abelian Higgs vortices in the near
Bogomolny regime.
\newblock \textsl{Commun. Math. Phys.} \textbf{159} (1994) 51--91


\bibitem{WhiWat}
\textsc{E.T. Whittaker, G.N. Watson}: {\it A Course of Modern Analysis}, 4th Edition, Cambridge University Press, 
1927

\bibitem{Wil}
\textsc{F. Wilczek}: {\it Fractional Statistics and Anyon Superconductivity}, World Scientific, 
1990


\bibitem{WitSMT}
\textsc{E. Witten}: Supersymmetry and Morse theory.
\newblock \textsl{J. Diff. Geom.} \textbf{17} (1982) 661--692


\bibitem{Wit-QFS} \textsc{E. Witten}: {Dynamics of quantum field theory} (Notes by P.~Etingof, L.~Jeffrey, D.~Kazhdan, J.~Morgan and D.~Morrison).
In: {\sc P.~Deligne, P.~Etingof, D.S.~Freed,
  L.C.~Jeffrey, D.~Kazhdan, J.W.~Morgan, D.S.~Morrison, E.~Witten} (Eds.):
   {\it Quantum Fields and Strings: A Course for Mathematicians}, vol.~2, \newblock American
  Mathematical Society, 1999


\end{small}

\end{thebibliography}

\end{document}